\PassOptionsToPackage{url}{hyphens}
\PassOptionsToPackage{hyperref}{hidelinks=true,breaklinks=true}
\documentclass[10pt, pageno,nohyperref]{jpaper}

\usepackage{adjustbox}
\usepackage{siunitx}
\usepackage{multirow}
\usepackage[T1]{fontenc}
\usepackage[tt=false, type1=true]{libertine}
\usepackage[varqu]{zi4}
\usepackage[libertine]{newtxmath}
\usepackage{tabularx}

\usepackage{fancyhdr}
\usepackage{pifont} 
\usepackage[normalem]{ulem}
\usepackage[sort,nocompress]{cite}
\usepackage{authblk}
\usepackage{xstring}
\usepackage[dvipsnames]{xcolor}
\usepackage{enumitem}
\usepackage[italic]{mathastext}
\usepackage[hang,bottom]{footmisc}
\usepackage[all]{nowidow}
\usepackage{bm}
\usepackage{dblfloatfix}
\usepackage{pbalance}

\usepackage{tikz}
\usepackage[hidelinks,breaklinks=true]{hyperref}

\DeclareSIUnit{\nothing}{\relax}

\begin{document}

\title{Leveraging ECRAM for Edge Continual Learning\thanks{Extended abstract of poster presented at 2025 IBM IEEE CAS/EDS AI Compute Symposium (AICS).}}

\author{Nabila Tasnim\qquad\qquad Haoran Liu\qquad\qquad Qing Cao\qquad\qquad Saugata Ghose}
\affil{
   \vspace{-5pt}
   University of Illinois Urbana-Champaign
}

\date{}

\newif\ifcameraready
\camerareadyfalse

\newif\ifonlineversion
\onlineversionfalse

\ifcameraready
  \newcommand{\todo}[1][]{}
  \newcommand{\sg}[1]{{#1}}
\else
  \newcommand{\todo}[1][]{\textbf{\scriptsize \fcolorbox{black}{red}{\color{white}{TODO}}} \underline{$\overline{\hbox{\emph{#1}}}$}}
  \newcommand{\sg}[1]{\textcolor{Red}{#1}}
\fi

\makeatletter
\renewcommand\paragraph[1]{\vspace{5pt}\noindent\textbf{#1}.}
\makeatother

\renewcommand{\citepunct}{,\penalty\citepunctpenalty\,}
\renewcommand{\citedash}{--}

\newcommand{\circled}[1]{%
  \tikz[baseline=(char.base)]{
  \node[shape=circle,draw,inner sep=0.5pt,fill=black,text=white,font=\small\bfseries] (char) {#1};}%
}%

\sisetup{detect-all = true, range-phrase = {--}, range-units=single, per-mode=symbol, separate-uncertainty=true, multi-part-units=single}

\maketitle

\pagenumbering{arabic}

\section{Introduction}
\label{sec:intro}

Several edge computing platforms, such as autonomous vehicles and smart sensing devices, need to adapt to dynamic environments in real time by learning from new data in the field.
One way to achieve this is to leverage \emph{machine learning} (ML), by retraining models at runtime to incorporate recently sensed data.
Unfortunately, this is not possible with conventional ML approaches: such retraining needs to ensure that the data previously trained on is not forgotten, but edge devices would need to spend significant storage resources and energy to store the old data and repeatedly perform full model training~\cite{shaheen2022continual, lesort2020continual, ullah2009you}.
\emph{Continual learning} has emerged as a promising solution, by incorporating techniques that successfully combine a highly summarized version of previously trained data (to avoid catastrophic forgetting) with recently sensed data~\cite{lesort2020continual}.
As is the case with other ML algorithms, continual learning generates significant data movement between general-purpose CPUs/GPUs and memory, which wastes energy and latency~\cite{mutlu2022modern, dally2022model}.

\emph{In-memory computing} (IMC; also known as \emph{processing-using-memory} or PUM)~\cite{10019367, kang2020deep, sebastian2020memory, ielmini2018memory} can curtail this waste and make continual learning feasible at the edge, but it faces two unique challenges that prior IMC-for-ML works did not face. First, \emph{IMC architectures make use of noisy computation operations that significantly harm training accuracy}.
The analog processing performed during IMC is highly sensitive to circuit and device non-idealities~\cite{5389202}. 
Existing works accept the reduced accuracy as a trade-off to improve energy efficiency, throughput, and/or area density~\cite{merolla2014million}. 
Unfortunately, given the limited input data used by continual learning algorithms, and the closed-loop nature of training algorithms, even minimal computing errors in gradient updates can accumulate and magnify over training epochs, resulting in significant training inaccuracies~\cite{xi2020memory, yu2018neuro}. Second, \emph{IMC architectures have poor and often incomplete support for resource-efficient training}.
Most prior works (e.g., \cite{sebastian2020memory, shafiee2016isaac}) focus on the use of multi-bit resistive non-volatile memories (NVMs) to improve inference, a read-only operation that does not update the ML model.
While a few works explore in-memory training~\cite{ortner2025rapid, rasch2024fast}, they focus on the NVM array design or circuit mechanisms, leaving out critical details about system-level integration. Existing works on IMC for continual learning \cite{zhang2024efficient, zhang2024device, liu2024edge} do not address one or both of these challenges.

\section{IMC Acceleration of Continual Learning}
\label{sec:keyideas}

We introduce \emph{CLASP} (the \emph{Continual Learning Acceleration System Platform}), which to our knowledge is the first end-to-end system with IMC acceleration for continual learning.
The hardware and software of CLASP are co-designed to support a wide range of continual learning algorithms, through software-visible assembly-level instructions that can be incorporated without constraints into ML-based algorithms.
We address the challenge of analog non-idealities by using electrochemical RAM (ECRAM)~\cite{cui2023cmos} as our in-memory substrate.

\subsection{End-to-End Overview of CLASP}
\label{sec:keyideas:overview}

The main source of analog non-idealities in existing IMC accelerators is the inherent properties of the devices selected.
These prior works typically make use of ReRAM, PCM, STT-MRAM, or SOT-MRAM, which exhibit limited conductance state support, non-linearity, low precision, and programming asymmetry, all of which make these device technologies ill-suited for in-memory training~\cite{ielmini2018memory, stern2021sub}.
CLASP is instead designed around a \emph{back-end-of-line} (BEOL) compatible ECRAM device that can overcome these challenges.
To verify the suitability of ECRAM for CLASP, we fabricated an improved version of our previous ECRAM array~\cite{cui2023cmos}.
The array exhibits symmetric programming characteristics, high modulation speed, and the ability to reliably represent 256~different conductance states with \textless 1\% cycle-to-cycle and device-to-device variability.

\begin{figure*}[t]
    \centering%
    \includegraphics[width=0.85\linewidth]{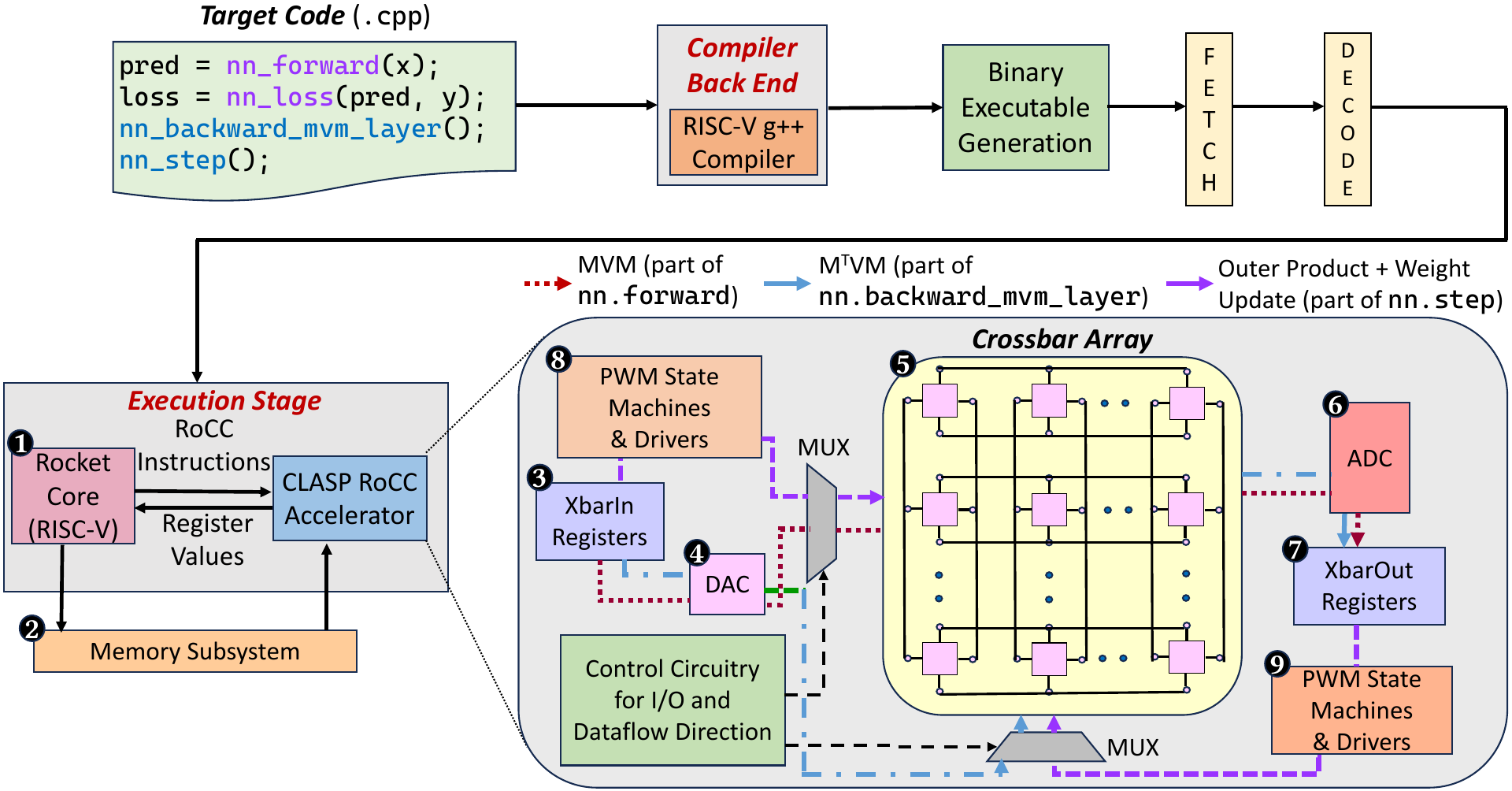}%
    \caption{Overview of CLASP.}%
    \label{fig:architecture}%
\end{figure*}

Figure~\ref{fig:architecture} shows a high-level overview of CLASP.
At its core is an IMC accelerator consisting of multiple ECRAM crossbar tiles (i.e., arrays), each with BEOL-integrated peripheral circuits (e.g., analog-to-digital converters, digital-to-analog converters), and with the tiles connected via an on-chip network. Within a crossbar, each weight of a neural network $w_{ij}$ is encoded as the programmable conductance $G_{ij}$ of an ECRAM device at row~$j$, column~$i$. Inputs are applied as wordline voltages $V_j$, and the bitline (column) currents $\left(I_i = \sum_{j} G_{ij} V_j \right)$ are sensed as outputs~\cite{xiao2020analog}.

The IMC accelerator is integrated as a tightly coupled Rocket custom coprocessor (RoCC) to an out-of-order RISC-V CPU core~\cite{asanovic2016rocket}, and makes use of 2.5D integration.
Such tight integration allows the IMC accelerator to have access to the CPU's memory system, allowing for training intermediates to rely in the CPU's caches without the need for DMA engines and additional scratchpads/caches.
As a result, CLASP uses two modes to communicate with the CPU core.
First, IMC commands (implemented as custom RISC-V instructions), responses, and scalar register exchanges are performed over the RoCC interface.
Second, bulk data transfers during IMC command execution are performed through the CPU core's cache-coherent memory interface.
By combining the RoCC interface with cache-coherent bulk data accesses, we can reuse existing RISC-V compiler back-ends to generate continual learning binaries.

\subsection{Hardware--Software Co-Design}
\label{sec:keyideas:codesign}

\emph{Neural network} (NN) training or continual learning algorithms are sensitive to the order of operations. For example, to execute simple NN training with analog IMC, we need to account for three tiers of dependencies: (1)~inter-stage ordering (forward $\rightarrow$ backward $\rightarrow$ optimizer), (2)~inter-layer sequencing, (3)~intra-layer micro-stages (DAC $\rightarrow$ MVM $\rightarrow$ ADC, loss and gradient computation, reverse-path MVM, weight update with pulses). CLASP must carefully consider all of these dependencies to avoid making use of stale results. To this end, we introduce a training-aware scheduler for CLASP that tackles this problem by maintaining strict serialization between dependent operations of all stages.  

Furthermore, we need to ensure correct operand selection and dataflow direction across different stages. For example, the backward stage operation flows in the reverse direction compared tothe  forward stage, while the optimizer stage flows in the same direction as the forward stage. In addition, each stage requires a different set of operands. To support these varying requirements, we introduce lightweight control circuitry (muxes, bidirectional interconnect) that dynamically selects the appropriate operand sources and configures the dataflow direction for each execution stage.

Our fabricated ECRAM device can support 8~bits (256 distinguishable states) of data per device. To match with this, we adopt 8-bit quantization for weights, layer inputs, and outputs. To execute continual learning algorithms with our system, we use a \emph{rectified linear unit} (ReLU) as the activation function for all hidden layers, while the output layer uses a task-specific activation function. ReLU produces sparse activations by setting negative values to zero. When an activation output is zero, the corresponding operation in the next layer can be skipped. This avoids unnecessary DAC driving and reduces the number of multiply--accumulate and ADC operations. This algorithm design choice reduces both overall latency and energy.

Our system exposes a custom set of RISC-V instructions for on-device learning: \texttt{nn\_init}, \texttt{nn\_load\_input}, \texttt{nn\_forward}, per-layer \texttt{nn\_backward\_mvm\_layer}, and \texttt{nn\_optimizer\_step}. These instructions provide a common interface that can support different learning algorithms without requiring any hardware modifications. CLASP requires only changing the CPU-side code and invoking the CLASP scheduler to orchestrate the appropriate sequence of instructions for different algorithms. For example, the following learning algorithms can be supported:

\begin{itemize}[leftmargin=*]
  \item \textbf{Transfer Learning:} Transfer learning adapts a pre-trained model to a new task by updating only a subset of the network. It can be implemented by invoking \texttt{nn\_forward} for inference, skipping \texttt{nn\_backward\_mvm\_layer} for the frozen layers (or setting their gradients to zero), and calling \texttt{nn\_optimizer\_step} only for the trainable layers. 
  
  \item \textbf{Few-Shot (Episodic) Learning:} Few-shot learning enables the model to learn a new task from only a small number of labeled examples. The model repeatedly performs forward propagation, computes the loss and gradients on the CPU, executes \texttt{nn\_backward\_mvm\_layer} to propagate gradients through the trainable layers, and finally updates the weights using \texttt{nn\_optimizer\_step}. This sequence is repeated for the support set before evaluating the adapted model on the query set.
  
  \item \textbf{Self-Supervised (Contrastive/Distillation) Learning:} Self-supervised learning trains a model without manual labels by generating supervisory signals from the input data itself. For example, two augmented versions of the same image are passed through the network, and a contrastive or distillation loss is computed on the CPU. The resulting gradients are then propagated using \texttt{nn\_backward\_mvm\_layer}, followed by \texttt{nn\_optimizer\_step} to update the model parameters. The same instruction sequence is reused for each augmented view.
  
\end{itemize}

\subsection{CLASP Execution Flow Example}
\label{sec:keyideas:flow}

\begin{figure}[h]
    \centering%
    \includegraphics[width=\linewidth]{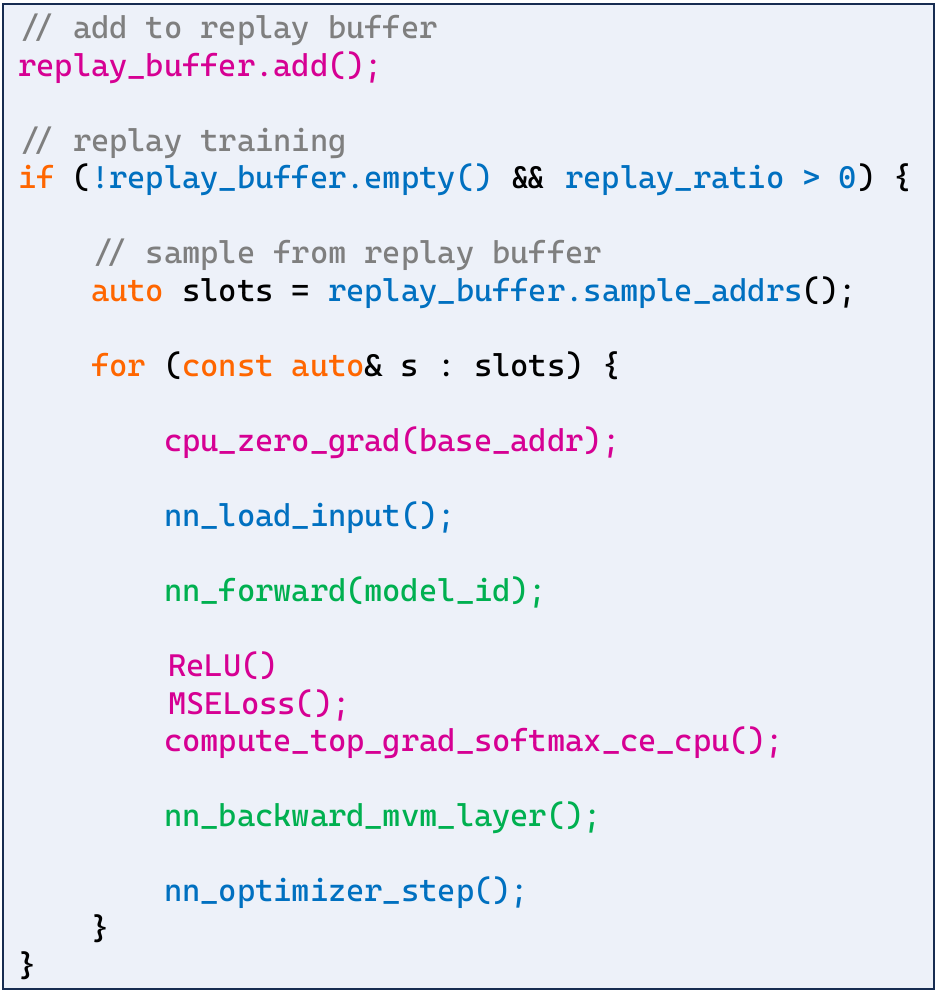}%
    \caption{Experience replay algorithm implemented with CLASP pseudocode.}%
    \label{fig:replay}%
\end{figure}

To illustrate how CLASP operates during continual learning, we walk through a code-to-hardware example of running \emph{experience replay}~\cite{replay}, a popular continual learning algorithm.
Experience replay stores past data information in a reservoir-sampled buffer and performs training with both the newly generated data and the stored past-data samples. 
Figure~\ref{fig:replay} shows the pseudocode for experience replay, written using the instructions that we introduce for CLASP control in Section~\ref{sec:keyideas:codesign}. 
The experience replay algorithm stores past information in a buffer and reuses this data during training to prevent forgetting.  
We annotate steps that walk through the execution flow of experience replay in Figure~\ref{fig:architecture}.

The RISC-V CPU (\circled{1} in Figure~\ref{fig:architecture}) fetches and decodes the instructions from the program (which we write using C++ and compile with a RISC-V version of \texttt{g++}).
At the start of the experience replay algorithm, whenever new data arrives that needs to be incorporated in the model, the data is compared against existing data samples in the buffer.
If the new contains enough new information over the existing samples, then it is added into the buffer (\texttt{replay\_buffer.add()}). After that, when the RISC-V core (\circled{1} in Figure~\ref{fig:architecture}) encounters any of our custom instructions, it offloads that operation to the accelerator. 
Next, upon issuing a \texttt{nn\_load\_input} instruction,
the CPU commands the IMC accelerator to load input data from memory (\circled{2}) into input registers (XBarIn; \circled{3}). When the core issues a \texttt{nn\_forward} instruction, it commands the accelerator to pass the input register data through the DAC (\circled{4}) into the crossbar bitlines (\circled{5}), performing a layer calculation.
The crossbar outputs traverse through the ADCs and are stored in XBarOut registers (\circled{6}); this data can be sent back to the RISC-V core to compute non-linear activations. The red dotted arrow shows the path for \texttt{nn\_forward}.

For training, to compute the loss, derivative of loss ($\delta$), and activation (e.g., ReLU, MSELoss steps of Figure~\ref{fig:replay}), the RISC-V core loads the activation outputs from memory, performs these computations, and stores the results back to memory.
To compute the layer gradient $W^\top \delta$ as part of the \texttt{nn\_backward\_mvm\_layer} instruction, $\delta$ is loaded into XBarIn (\circled{3}) and follows the path indicated by the light blue dash--dot arrow in Figure~\ref{fig:architecture} (\circled{5}--\circled{7}).
To update the weights via analog stochastic gradient descent (\texttt{nn\_optimizer\_step}), $\delta$ is loaded into XBarOut (\circled{7}), and the inputs are loaded into XBarIn (\circled{3}) and then passed to pulse width modulation (PWM) drivers  (\circled{8} and  \circled{9}, respectively) to generate the corresponding update pulses.
These pulses are fed to the crossbar array (\circled{5}) via the purple dashed arrow in Figure~\ref{fig:architecture}.

\section{Evaluation}

To model CLASP end-to-end, we integrate the Structural Simulation Toolkit (SST)~\cite{sst.website} with the IBM Analog Hardware Acceleration Kit (AIHWKIT)~\cite{aihwkit}.
Our IMC accelerator is modeled in AIHWKIT, where the our fabricated ECRAM device characteristics are used to define the underlying hardware behavior.
We add AIHWKIT as a coprocessor in SST's Vanadis module~\cite{vanadis.sst}.
Vanadis models the RISC-V core, and we use other components of SST to faithfully model the memory system, interconnect, and OS.

In order to capture the impact of our ECRAM device, we also demonstrate variants of CLASP where the ECRAM is replaced by ReRAM and PCM.
As ReRAM and PCM devices are unable to accurately represent 8-bit values, we employ bit-slicing, and evaluate ReRAM devices capable of 2-bit and 4-bit representations~\cite{gong2022deep}, and PCM devices capable of 2-bit representations~\cite{li2023optimization}.

To demonstrate CLASP's flexibility, we evaluate it using two continual learning algorithms: 
(1)~\emph{learning without forgetting} (LwF)~\cite{li2017learning} and 
(2)~experience replay~\cite{replay}. 
LwF remembers old task information by distilling the  previous task model’s frozen \emph{logits} (i.e.,  the non-normalized numerical scores out by the final layer) while learning the new task data. We implement the  algorithms in PyTorch using a \emph{multi-layer perceptron} (MLP) model (model size: 784 $\rightarrow$ 128 $\rightarrow$ 10), allowing us to compare equivalent software models across all of our evaluated systems. The results are generated with the MNIST dataset~\cite{6296535} as input. We use an NVIDIA GeForce RTX 4090 GPU~\cite{nvidiaNVIDIAGeForce} as our baseline.

Table~\ref{accuracy_comparison} shows the output accuracy for each of our configurations as we train on 60k~images from the MNIST dataset, with a separate 10k-image test set. 
We make two observations from the table.
First, CLASP with ECRAM has significantly higher accuracy compared to CLASP using other emerging memory devices. Both the ReRAM- and PCM-based CLASP have transfer characteristics that are highly non-linear and asymmetric, making gradient computations error prone.
We note that the accuracy drops more for 2-bit ReRAM than 4-bit ReRAM, as
even though the 2-bit representation generates fewer read errors per device, we must employ larger arrays to hold the entire model, which contributes more noise~\cite{bhattacharjee2022examining}.
Second, we observe that CLASP with 8-bit ECRAM achieves accuracies that approach our baseline in-GPU training.

\begin{table}[h]
  \centering\footnotesize
  \vspace{-12pt}
  \caption{Accuracy of learning algorithms with CLASP.}
  \label{accuracy_comparison}
  \begin{tabular}{|c||c|c|}
    \hline
    \textbf{Method} & \textbf{LwF Accuracy} & \textbf{Replay Accuracy} \\
    \hline
    \hline
    GPU w/ PyTorch & 54.53\% & 90.91\% \\
    \hline
    CLASP: 2-bit ReRAM & 35.20\% & 40.27\% \\
    CLASP: 4-bit ReRAM & 40.23\% & 59.80\% \\
    CLASP: 2-bit PCM & 17.18\% & 28.25\% \\
    CLASP: 2-bit ECRAM & 41.66\% & 62.77\% \\
    CLASP: 4-bit ECRAM & 48.80\% & 72.48\% \\
    CLASP: 8-bit ECRAM & 51.43\% & 84.20\% \\
    \hline
  \end{tabular}
\end{table}

\begin{figure}[h]
    \centering%
    \includegraphics[width=\linewidth]{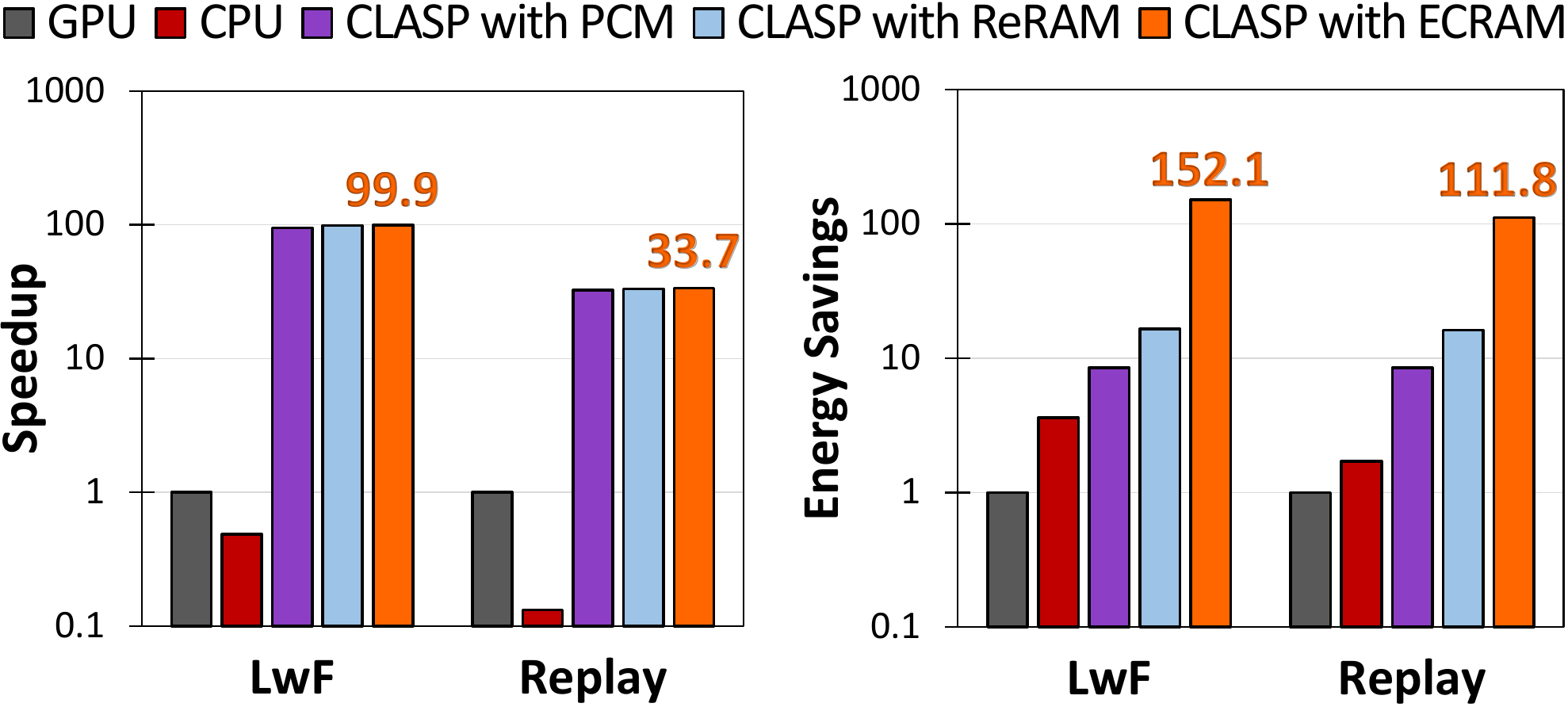}%
    \caption{Speedup (left) and energy savings (right) for CLASP normalized to GPU; y-axis in log scale.}%
    \label{fig:results}%
\end{figure}

Figure~\ref{fig:results} shows the speedup (left) and energy savings (right) for different variants of CLASP, normalized to our GPU baseline, using a 1k-image training set sample from MNIST.
We include a comparison to CPU-only training using a \SI{2.3}{\giga\hertz} RISC-V CPU. 
We make two observations from the figure.
First, CLASP with ECRAM achieves 100$\times$ and 34$\times$ speedup over the GPU baseline for LwF and experience replay, respectively. 
We note similar speedups across all memory types, as the in-CPU gradient processing and intermediate data storage become the training pipeline bottlenecks after IMC acceleration, making the specific performance of the device a second-order factor.
Second, we note that the GPU performance improvement compared to CPU-only training is lower than expected as we are using small model sizes, for which GPU launch overheads dominate.
Third, CLASP with ECRAM achieves 152$\times$ and 112$\times$ energy savings over the GPU for LwF and experience replay, respectively, reflecting a combination of reduced data movement and low-energy multiplication and gradient calculation.

\section*{Acknowledgments}
We thank Shomik Chatterjee, Shaloo Rakheja, Divake Kumar, and Amit Ranjan Trivedi for their discussions on the project.
This work was supported in part by NSF grant CCF-2329096.

\bibliographystyle{IEEEtranS}
\bibliography{ref}

\begin{thebibliography}{10}
\providecommand{\url}[1]{#1}
\csname url@samestyle\endcsname
\providecommand{\newblock}{\relax}
\providecommand{\bibinfo}[2]{#2}
\providecommand{\BIBentrySTDinterwordspacing}{\spaceskip=0pt\relax}
\providecommand{\BIBentryALTinterwordstretchfactor}{4}
\providecommand{\BIBentryALTinterwordspacing}{\spaceskip=\fontdimen2\font plus
\BIBentryALTinterwordstretchfactor\fontdimen3\font minus \fontdimen4\font\relax}
\providecommand{\BIBforeignlanguage}[2]{{%
\expandafter\ifx\csname l@#1\endcsname\relax
\typeout{** WARNING: IEEEtranS.bst: No hyphenation pattern has been}%
\typeout{** loaded for the language `#1'. Using the pattern for}%
\typeout{** the default language instead.}%
\else
\language=\csname l@#1\endcsname
\fi
#2}}
\providecommand{\BIBdecl}{\relax}
\BIBdecl

\bibitem{asanovic2016rocket}
K.~Asanovi{\'c}, R.~Avi{\v{z}}ienis, J.~Bachrach, S.~Beamer, D.~Biancolin, C.~Celio, H.~Cook, D.~Dabbelt, J.~Hauser, A.~Izraelevitz, S.~Karandikar, B.~Keller, D.~Kim, J.~Koenig, Y.~Lee, E.~Love, M.~Maas, A.~Magyar, H.~Mao, M.~Moret{\'o}, A.~Ou, D.~A. Patterson, B.~Richards, C.~Schmidt, S.~Twigg, H.~Vo, and A.~Waterman, ``{The Rocket Chip Generator},'' {Univ.\ of California, Berkeley}, Tech. Rep. UCB/EECS-2016-17, 2016.

\bibitem{bhattacharjee2022examining}
A.~Bhattacharjee, L.~Bhatnagar, and P.~Panda, ``{Examining and Mitigating the Impact of Crossbar Non-Idealities for Accurate Implementation of Sparse Deep Neural Networks},'' in \emph{DATE}, 2022.

\bibitem{cui2023cmos}
J.~Cui, F.~An, J.~Qian, Y.~Wu, L.~L. Sloan, S.~Pidaparthy, J.-M. Zuo, and Q.~Cao, ``{CMOS-Compatible Electrochemical Synaptic Transistor Arrays for Deep Learning Accelerators},'' \emph{Nature Electronics}, Apr. 2023.

\bibitem{dally2022model}
W.~J. Dally, ``{On the Model of Computation: Point},'' \emph{CACM}, Sep. 2022.

\bibitem{6296535}
L.~Deng, ``{The MNIST Database of Handwritten Digit Images for Machine Learning Research [Best of the Web]},'' \emph{IEEE SPM}, Nov. 2012.

\bibitem{10019367}
D.~Fick, ``{Analog Compute-in-Memory for AI Edge Inference},'' in \emph{IEDM}, 2022.

\bibitem{gong2022deep}
N.~Gong, M.~J. Rasch, S.-C. Seo, A.~Gasasira, P.~M. Solomon, V.~Bragaglia, S.~Consiglio, H.~Higuchi, C.~Park, K.~Brew, P.~C. Jamison, C.~Catano, I.~Saraf, F.~F. Athena, C.~Silvestre, X.~Liu, B.~Khan, N.~Jain, S.~McDermott, R.~Johnson, I.~C. Estrada-Raygoza, J.~Li, T.~Gokmen, N.~Li, R.~Pujari, F.~Carta, H.~Miyazoe, M.~M. Frank, D.~Koty, Q.~Yang, R.~D. Clark, K.~Tapily, C.~S. Wajda, A.~Mosden, J.~Shearer, A.~W. Metz, S.~Teehan, N.~Saulnier, B.~J. Offrein, T.~Tsunomura, G.~J. Leusink, V.~Narayanan, and T.~Ando, ``{Deep Learning Acceleration in \qty{14}{\nano\metre} CMOS Compatible ReRAM Array: Device, Material and Algorithm Co-Optimization},'' in \emph{IEDM}, 2022.

\bibitem{ielmini2018memory}
D.~Ielmini and H.-S.~P. Wong, ``{In-Memory Computing With Resistive Switching Devices},'' \emph{Nature Electronics}, Jun. 2018.

\bibitem{kang2020deep}
M.~Kang, S.~K. Gonugondla, and N.~R. Shanbhag, \emph{{Deep In-Memory Architectures for Machine Learning}}.\hskip 1em plus 0.5em minus 0.4em\relax Springer, 2020.

\bibitem{lesort2020continual}
T.~Lesort, V.~Lomonaco, A.~Stoian, D.~Maltoni, D.~Filliat, and N.~D{\'i}az-Rodr{\'i}guez, ``{Continual Learning for Robotics: Definition, Framework, Learning Strategies, Opportunities and Challenges},'' \emph{Information Fusion}, Jun. 2020.

\bibitem{li2023optimization}
N.~Li, C.~Mackin, A.~Chen, K.~Brew, T.~Philip, A.~Simon, I.~Saraf, J.-P. Han, S.~G. Sarwat, G.~W. Burr, M.~J. Rasch, A.~Sebastian, V.~Narayanan, and N.~Saulnier, ``{Optimization of Projected Phase Change Memory for Analog In-Memory Computing Inference},'' \emph{Advanced Electronic Materials}, Jun. 2023.

\bibitem{li2017learning}
Z.~Li and D.~Hoiem, ``{Learning Without Forgetting},'' \emph{TPAMI}, Dec. 2018.

\bibitem{liu2024edge}
S.~Liu, R.~M. Radway, X.~Wang, F.~Moro, J.-F. Nodin, K.~Jana, S.~Du, L.~R. Upton, W.-C. Chen, J.~Chen, H.~Li, F.~Andrieu, E.~Vianello, P.~Raina, S.~Mitra, and H.-S.~P. Wong, ``{Edge Continual Training and Inference With RRAM-Gain Cell Memory Integrated on Si CMOS},'' in \emph{IEDM}, 2024.

\bibitem{merolla2014million}
P.~A. Merolla, J.~V. Arthur, R.~Alvarez-Icaza, A.~S. Cassidy, J.~Sawada, F.~Akopyan, B.~L. Jackson, N.~Imam, C.~Guo, Y.~Nakamura, B.~Brezzo, I.~Vo, S.~K. Esser, R.~Appuswamy, B.~Taba, A.~Amir, M.~D. Flickner, W.~P. Risk, R.~Manohar, and D.~S. Modha, ``{A Million Spiking-Neuron Integrated Circuit With a Scalable Communication Network and Interface},'' \emph{Science}, Aug. 2014.

\bibitem{mutlu2022modern}
O.~Mutlu, S.~Ghose, J.~G{\'o}mez-Luna, and R.~Ausavarungnirun, ``{A Modern Primer on Processing in Memory},'' in \emph{Emerging Computing: From Devices to Systems: Looking Beyond Moore and Von Neumann}, M.~M. {Sabry Aly} and A.~Gupta, Eds.\hskip 1em plus 0.5em minus 0.4em\relax Springer, 2022.

\bibitem{vanadis.sst}
{National Technology \& Engineering Solutions of Sandia LLC}, ``{vanadis --- Introduction},'' \url{https://sst-simulator.org/sst-docs/docs/elements/vanadis/intro}.

\bibitem{nvidiaNVIDIAGeForce}
{NVIDIA Corp.}, ``{GeForce RTX 4090 Graphics Cards},'' \url{https://www.nvidia.com/en-us/geforce/graphics-cards/40-series/rtx-4090/}.

\bibitem{ortner2025rapid}
T.~Ortner, H.~Petschenig, A.~Vasilopoulos, R.~Renner, {\v{S}}.~Brglez, T.~Limbacher, E.~Pi{\~n}ero, A.~Linares-Barranco, A.~Pantazi, and R.~Legenstein, ``{Rapid Learning With Phase-Change Memory-Based In-Memory Computing Through Learning-to-Learn},'' \emph{Nature Communications}, Feb. 2025.

\bibitem{rasch2024fast}
M.~J. Rasch, F.~Carta, O.~Fagbohungbe, and T.~Gokmen, ``{Fast and Robust Analog In-Memory Deep Neural Network Training},'' \emph{Nature Communications}, Aug. 2024.

\bibitem{aihwkit}
M.~J. Rasch, D.~Moreda, T.~Gokmen, M.~{Le Gallo}, F.~Carta, C.~Goldberg, K.~{El Maghraoui}, A.~Sebastian, and V.~Narayanan, ``{A Flexible and Fast PyTorch Toolkit for Simulating Training and Inference on Analog Crossbar Arrays},'' in \emph{AICAS}, 2021.

\bibitem{replay}
D.~Rolnick, A.~Ahuja, J.~Schwarz, T.~P. Lillicrap, and G.~Wayne, ``{Experience Replay for Continual Learning},'' in \emph{NeurIPS}, 2019.

\bibitem{5389202}
A.~L. Samuel, ``{Some Studies in Machine Learning Using the Game of Checkers},'' \emph{IBM JRD}, Jan. 2000.

\bibitem{sst.website}
{Sandia National Laboratories}, ``{The Structural Simulation Toolkit},'' \url{https://sst-simulator.org/}.

\bibitem{sebastian2020memory}
A.~Sebastian, M.~{Le Gallo}, R.~Khaddam-Aljameh, and E.~Eleftheriou, ``{Memory Devices and Applications for In-Memory Computing},'' \emph{Nature Nanotechnology}, Jul. 2020.

\bibitem{shafiee2016isaac}
A.~Shafiee, A.~Nag, N.~Muralimanohar, R.~Balasubramonian, J.~P. Strachan, M.~Hu, R.~S. Williams, and V.~Srikumar, ``{ISAAC: A Convolutional Neural Network Accelerator With In-Situ Analog Arithmetic in Crossbars},'' in \emph{ISCA}, 2016.

\bibitem{shaheen2022continual}
K.~Shaheen, M.~A. Hanif, O.~Hasan, and M.~Shafique, ``{Continual Learning for Real-World Autonomous Systems: Algorithms, Challenges and Frameworks},'' \emph{JIRS}, Jan. 2022.

\bibitem{stern2021sub}
K.~Stern, N.~Wainstein, Y.~Keller, C.~M. Neumann, E.~Pop, S.~Kvatinsky, and E.~Yalon, ``{Sub-Nanosecond Pulses Enable Partial Reset for Analog Phase Change Memory},'' \emph{EDL}, Sep. 2021.

\bibitem{ullah2009you}
M.~M. Ullah, F.~Orabona, and B.~Caputo, ``{You Live, You Learn, You Forget: Continuous Learning of Visual Places With a Forgetting Mechanism},'' in \emph{IROS}, 2009.

\bibitem{xi2020memory}
Y.~Xi, B.~Gao, J.~Tang, A.~Chen, M.-F. Chang, X.~S. Hu, J.~{Van der Spiegel}, H.~Qian, and H.~Wu, ``{In-Memory Learning With Analog Resistive Switching Memory: A Review and Perspective},'' \emph{Proc.\ IEEE}, Jan. 2021.

\bibitem{xiao2020analog}
T.~P. Xiao, C.~H. Bennett, B.~Feinberg, S.~Agarwal, and M.~J. Marinella, ``{Analog Architectures for Neural Network Acceleration Based on Non-Volatile Memory},'' \emph{Applied Physics Reviews}, Sep. 2020.

\bibitem{yu2018neuro}
S.~Yu, ``{Neuro-Inspired Computing With Emerging Nonvolatile Memorys},'' \emph{Proc.\ IEEE}, Feb. 2018.

\bibitem{zhang2024device}
F.~Zhang, A.~Sridharan, W.~Hwang, F.~Xue, W.~Tsai, S.~X. Wang, and D.~Fan, ``{On-Device Continual Learning With STT-Assisted-SOT MRAM-Based In-Memory Computing},'' \emph{TCAD}, Aug. 2024.

\bibitem{zhang2024efficient}
F.~Zhang, A.~Sridharan, W.~Tsai, Y.~Chen, S.~X. Wang, and D.~Fan, ``{Efficient Memory Integration: MRAM-SRAM Hybrid Accelerator for Sparse On-Device Learning},'' in \emph{DAC}, 2024.

\end{thebibliography}

\end{document}